\documentclass[a4paper,12pt]{article}

\usepackage[sort&compress]{natbib}
\bibpunct{(}{)}{;}{a}{}{,} 

\usepackage{amsthm, amsmath, amssymb, mathrsfs, multirow}
\usepackage{graphicx} 
\usepackage{ifthen} 
\usepackage{amsfonts}
\usepackage[usenames,dvipsnames]{color}
\usepackage{epic}
\usepackage{url}
\usepackage{subfigure}
\usepackage{fullpage}

\theoremstyle{plain}
\newtheorem{thm}{\indent Theorem}

\theoremstyle{remark}

\theoremstyle{definition}

\theoremstyle{definition}

\theoremstyle{remark}
\newtheorem*{PRalgo}{\indent PR Algorithm}

\newcommand{\U}{\mathscr{U}}

\newcommand{\PP}{\mathbb{P}}

\newcommand{\RR}{\mathbb{R}}

\newcommand{\ith}{i^{\text{th}}}

\renewcommand{\phi}{\varphi}

\newcommand{\fdr}{\mathrm{fdr}}
\newcommand{\fdrhat}{\widehat{\mathrm{fdr}}}

\newcommand{\grad}{\nabla}

\DeclareMathOperator{\logit}{\mathrm{logit}}

\title{A nonparametric empirical Bayes framework for large-scale multiple testing}
\author{Ryan Martin \\ 
Department of Mathematics, Statistics, and Computer Science \\ 
University of Illinois at Chicago \\ 
\url{rgmartin@math.uic.edu} \\ 
\mbox{} \\ 
Surya T. Tokdar \\ 
Department of Statistical Science \\ 
Duke University \\ 
\url{tokdar@stat.duke.edu}}
\date{\today}

\begin{document}

\maketitle

\begin{abstract}
We propose a flexible and identifiable version of the two-groups model, motivated by hierarchical Bayes considerations, that features an empirical null and a semiparametric mixture model for the non-null cases.  We use a computationally efficient predictive recursion marginal likelihood procedure to estimate the model parameters, even the nonparametric mixing distribution.  This leads to a nonparametric empirical Bayes testing procedure, which we call PRtest, based on thresholding the estimated local false discovery rates.  Simulations and real-data examples demonstrate that, compared to existing approaches, PRtest's careful handling of the non-null density can give a much better fit in the tails of the mixture distribution which, in turn, can lead to more realistic conclusions.

\smallskip

\end{abstract}


\section{Introduction}
\label{S:intro}

Large-scale multiple testing problems arise in many applied fields such as genomics \citep{dudoitlaan2008,schaferstrimmer2005a}, proteomics \citep{dghosh2009}, astrophysics \citep{liang2004, miller2001}, and image analysis \citep{schwartzman2008, lindquist2008}, to name a few.  An abstract representation of the problem is testing a set of hypotheses 
\[ H_{0i}: \text{the $\ith$ case manifests a ``null'' behavior}, \quad i=1,\ldots,n \]
based on summary test statistics, or z-scores, $Z_1,\ldots,Z_n$. The null behavior of a single z-score $Z_i$ can be described by the ${\sf N}(0,1)$ distribution when $Z_i$ is defined as the Gaussian transform of a test statistic derived for the $\ith$ case, such as the two sample t-statistic comparing treatment to control. Although this characterization leads to a simple rejection rule for the $\ith$ case in isolation, it is found insufficient when all $n$ tests in are to be performed, particularly when $n$ is very large. In fact, one of the major developments of modern statistics has been the philosophical shift from treating the z-scores as mutually independent to treating them as exchangeable \citep{efrontibs}. Consequently, recent work on large-scale simultaneous testing has focused on Bayesian models and, in particular, empirical Bayes methods that allow for information sharing between cases, even though separate decisions will be made for each case.  

An elegant formalization of the large-scale simultaneous testing problem is the \emph{two-groups model} \citep{efron2004,efron2007,efron2008} which assumes $Z_1,\ldots,Z_n$ arise from a mixture density
\begin{equation}
\label{eq:two-groups}
f(z) = \pi f_0(z) + (1-\pi) f_1(z),
\end{equation}
with $f_0$ and $f_1$, respectively, describing the null and non-null distributions of the z-scores.  \citet{efron2004, efron2008} argues that, for a variety of reasons, the case-specific theoretical null distribution ${\sf N}(0,1)$ may not be an adequate choice for $f_0$, and a more appropriate choice is the so-called empirical null distribution ${\sf N}(\mu, \sigma^2)$, where $\mu$ and $\sigma$ are to be estimated from data.  

Following Efron's original treatment, various new methods have been proposed for fitting and drawing inference from the two groups model of z-scores \citep{jincai2007, muralidharan2009}.  These methods, together with related methodology based on p-values or t-scores \citep[e.g.,][]{benjamini-hochberg, storey2003}, have been widely used in biological studies with high-throughput data, in particular to identify genes responsible for a phenotypical behavior based on microarray analysis. The single-summary-per-case approach of these methods offers substantial computational advantage over other approaches to analyze such data, such as those based on high-dimensional classification techniques \citep{golub, leeplus2003}.

\begin{figure}
\begin{center}
\subfigure[Leukemia z-scores]{\scalebox{0.72}{\includegraphics{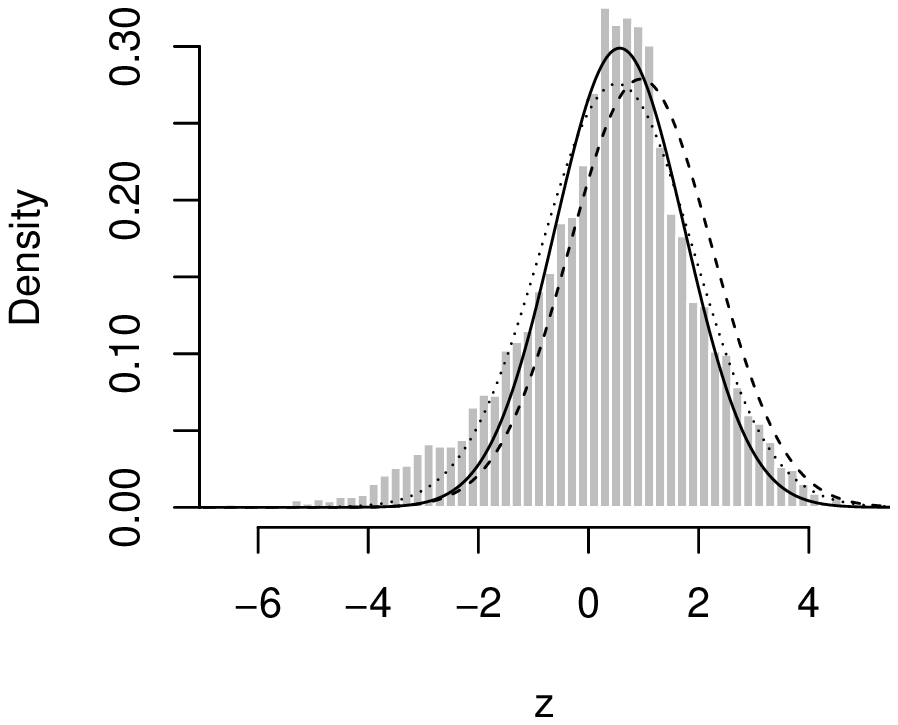}}} 
\subfigure[Breast cancer z-scores]{\scalebox{0.72}{\includegraphics{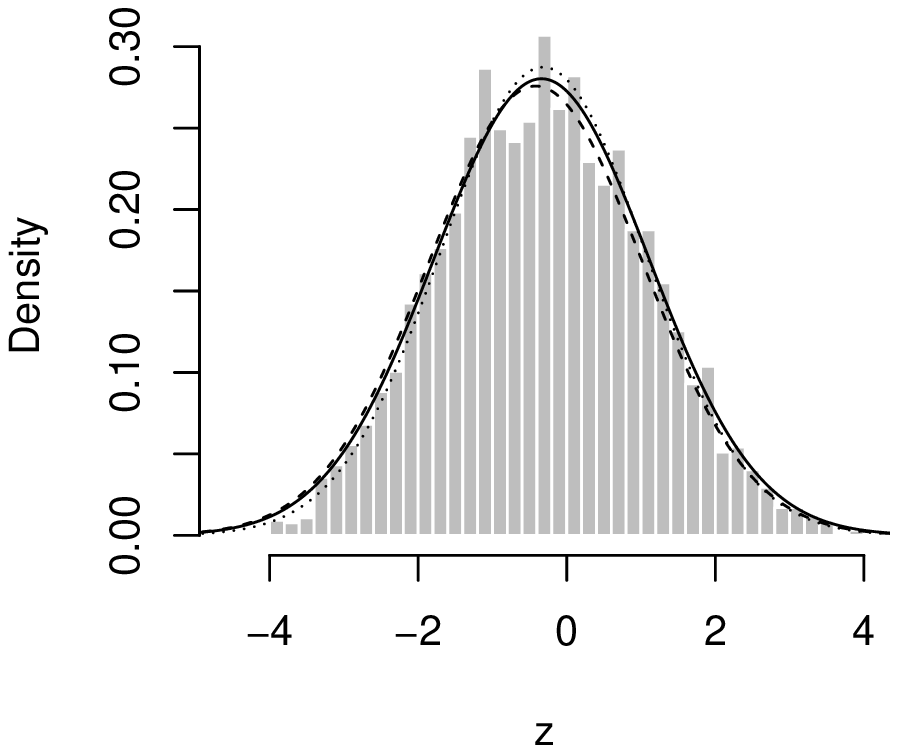}}}
\label{fig:badfit}
\caption{Density histogram of z-scores from leukemia microarray data \citep{golub} and breast cancer data \citep{hedenfalk2001} and estimates of $\pi f_0$ based on the methods of \citet{efron2004} (---), \citet{jincai2007}($-$$-$) and \citet{muralidharan2009} ($\cdots$). }
\end{center}
\end{figure}

However, currently available methods for fitting \eqref{eq:two-groups} do not take full advantage of the two-groups formulation.  Motivated by applications to microarray studies, where typically a very small fraction of genes are linked with the phenotype, existing two-groups methods take a conservative approach of encouraging estimates of $\pi$ close to 1.  While this is reasonable for many applications, there are scientific studies where such a conservative approach fails to detect any or a majority of the interesting cases.  Figure~\ref{fig:badfit} reports two such microarray studies, a leukemia study by \citet{golub} and a breast cancer study by \citet{hedenfalk2001}; more details are given in Section~\ref{S:examples}.  As shown in the figure, existing methods each produce estimates of the null component $\pi f_0$ that cover one or both tails of the z-score histogram, leaving little to be explained by the non-null component $(1-\pi)f_1$.  Consequently, zero discoveries of interesting genes are made in one or both tails; see Table~\ref{tab:real} in Section~\ref{S:examples}. High-dimensional classification-based analyses \citep{golub, hedenfalk2001, leeplus2003}, on the other hand, identify a number of interesting genes on either tail for each of the two studies.

In this paper we consider a new likelihood-based analysis of the two-groups model, with a regularization on $\mu, \sigma, \pi$ and a semiparametric specification of the non-null density $f_1$.  We employ a mixture representation of $f_1$ that gives it heavier tails than $f_0$ to reflect the belief that z-scores from the non-null cases are likely to be larger in magnitude than those from the null cases.  The null weight $\pi$ is given a beta prior with a center close to one but with a relatively long left tail.  Additionally, we use a prior on  $(\mu, \sigma)$ to reflect the belief that this vector is likely to be close to $(0,1)$. 

Compared to the existing methods based on z-scores, our proposal allows a wider range of estimates of $\pi$.  For scientific studies where the existing methods discover a fair number of interesting cases, our method makes similar discoveries.  But for other studies where existing methods fail, such as the two studies mentioned earlier, our method makes discoveries that are comparable to those found via high-dimensional classification methods.  A similar adaptability property manifests in a simulation study where z-scores are generated according to \eqref{eq:two-groups} with $\pi$ ranging between 0.75 to 0.99; see Table~\ref{tab:pi-est} in Section~\ref{S:simulations}.  

Despite a nonparametric specification of $f_1$ and a likelihood-based analysis, our treatment of the two-groups model retains the computational efficiency that is hallmark of methods based on z-scores.  This has been possible due to recent developments on a stochastic algorithm due to \citet{newton02} called {\it predictive recursion}, for estimation of mixing densities with respect to any arbitrary dominating measure; see also \citet{nqz}. Theoretical properties of this algorithm are addressed in \citet{ghoshtokdar}, \citet{martinghosh}, \citet{tmg}, and \citet{mt-rate}.  \citet{mt-prml} show how this algorithm can be used in a hierarchical mixture model to construct a likelihood function over non-mixing model parameters, marginalized over the mixing density. This marginal likelihood is shown to have strong connections to the marginal likelihood under a Bayesian Dirichlet process mixture model.  We adopt this marginal likelihood calculation to the two groups model, with $\mu$, $\sigma$, $\pi$ and a scaling parameter in the specification of $f_1$ serving as the non-mixing parameters.

For the multiple testing problem, we adopt the strategy of mimicking the Bayes oracle rule by thresholding a plug-in estimate of the local false discovery rate, similar to \citet{efron2004, efron2008}, \citet{jincai2007}, and \citet{muralidharan2009}.  Simulations presented in Section~\ref{S:simulations} show that the proposed method, called PRtest, is more adaptive to asymmetry in the non-null density $f_1$ and to the degree of sparsity characterized by $\pi$.  Performance of PRtest in an interesting example using the artificial microarray data of \citet{choe2005} is addressed in Section~\ref{S:examples}.  In this example, the set of interesting genes is known and we find that PRtest performs considerably better than existing methods and strikingly similar to the oracle.  Likewise, for the leukemia and hereditary breast cancer studies, we find that the PR-based estimation produces a better fit in the tails of the distribution than that seen in Figure~\ref{fig:badfit} and, consequently, we are able to identify a number of interesting genes in each example.  The identified genes are, in fact, consistent with those identified by more sophisticated high-dimensional classification-based techniques. 


\section{Model specification}
\label{S:model}

We take $f_0(z) = {\sf N}(z \mid \mu, \sigma^2)$, the normal density with unknown mean and variance $\mu$ and $\sigma^2$. The non-null density $f_1$ is taken to be a semiparametric mixture of the form
\begin{equation}
\label{eq:alternative}
f_1(z) = \int_{\U} {\sf N}(z \mid \mu + \tau \sigma u, \sigma^2) \psi(u) \,du,
\end{equation}
with $\psi$ a density with respect to the Lebesgue measure on $\U=[-1,1]$ and $\tau \geq 1$ a scaling factor.  An important consequence of the requirement that $\psi$ be a density is given in the following theorem; see Appendix~A for a proof.  

\begin{thm}
\label{thm:identifiable}
For $f_0$ and $f_1$ as described above, the parameters $(\mu, \sigma, \pi, \tau, \psi)$ in our version of the two-groups model are identifiable. 
\end{thm}

This result is useful because, in general, identifiability is not guaranteed for a two groups model \eqref{eq:two-groups} with an empirical null that involves unknown parameters. For our specification, the key to identifiability is the model feature that $f_1$, by virtue of averaging over locations shifts of $f_0$, has heavier tails than $f_0$.  This feature is scientifically relevant as it embeds the belief that z-scores in the tails of the histogram are more likely to correspond to non-null cases than null.  \citet{efron2008} incorporates a similar belief through a {\it zero-assumption}: most z-scores near zero are from the null component. However, such a zero-assumption can be too strong to allow learning from data and can lead to an estimate of $\pi f_0$ that has heavier tails than any reasonable histogram-smoothing estimate of $f$, as reported by \citet{strimmer2008} and illustrated in Figure~\ref{fig:badfit}. In comparison, separating $f_0$ and $f_1$ by their tails seems more practical; see Section~\ref{S:examples}.

An important feature of the model is that $f_0$ and the kernel being mixed in \eqref{eq:alternative} have a common scale.  This, along with the assumption that $\psi$ is a density, are the driving forces behind the characteristic function-based proof of Theorem~\ref{thm:identifiable}, not the distributional form of these densities.  In fact, the proof in Appendix~A applies when $f_0(z) = \sigma^{-1}p\bigl( (z - \mu) / \sigma \bigr)$ for a suitably smooth density $p$ symmetric about zero, and $f_1$ is likewise determined by $p$.  Here we focus only on the case where $p$ is the ${\sf N}(0,1)$ density, but other choices like a Student-t density with fixed degrees of freedom can also be entertained.

\section{Mixture models and predictive recursion}
\label{S:pr}

It is more convenient to write our specification of $f$ as the mixture model
\begin{equation}
\label{eq:mixture}
f(z) = \int_{\U} p(z \mid \theta, u) \, \Psi(du) 
\end{equation} 
with parameters $\theta = (\mu, \sigma, \tau)$, kernel $p(z \mid \theta, u) = {\sf N}(z \mid \mu + \tau\sigma u, \sigma^2)$, and mixing probability measure $\Psi$ on $\U$ that assigns a positive mass $\pi$ at $0 \in \U$ and distributes the remaining mass on $\U$ according to a Lebesgue density $\psi$. The collection of all such $\Psi$ is the set $\PP=\PP(\U,\nu)$ of probability measures that are absolutely continuous with respect to the measure $\nu$ defined as the sum of the Lebesgue measure on $\U$ and a point mass at 0.  The $\nu$-density of such an $\Psi$ will be denoted by $\pi\langle0\rangle + (1-\pi)\psi$.
 
Inference on $(\theta, \Psi)$ can be performed in a Bayesian setting with a prior distribution on $(\theta, \Psi)$.  A popular choice of prior distribution for the nonparametric probability measure $\Psi$ is the Dirichlet process prior \citep{ferguson1973}.  However, there are two practical difficulties in employing this inference framework for our model.  First, the Dirichlet process prior entertains only discrete probability measures, thus violating the important absolute continuity property of $\Psi$ with respect to $\nu$.  Second, despite recent advances in computing, fitting a Dirichlet process mixture model does not scale well with the number of observations $n$.  For microarray studies, $n$ ranges from thousands to tens of thousands, whereas for more recent single nucleotide polymorphism studies, $n$ can reach several hundreds of thousands.  For such massive data sets, fitting a Dirichlet process mixture model can be fairly time consuming, nullifying some of the advantages of the two-groups framework. 
 
As an alternative, we estimate $(\theta, \Psi)$ via the predictive recursion (PR) methodology \citep{newton02, mt-prml}.  PR is a stochastic algorithm for estimating a mixing distribution $\Psi$ in \eqref{eq:mixture} through fast, recursive updates that have a strong connection with posterior updates for Dirichlet process mixture models.  The algorithm accommodates user-specified absolute continuity constraints on the mixing distribution and enjoys attractive convergence properties under mild conditions with allowance for model misspecification \citep{ghoshtokdar, tmg, martinghosh, mt-rate}.  However, Newton's original proposal can estimate the mixing distribution only when the kernel being mixed is known exactly, i.e., for \eqref{eq:mixture}, an estimate of $\Psi$ is available only when $\theta$ is known.  To resolve this difficulty, \citet{mt-prml} introduce a ``marginal likelihood'' function for non-mixing parameters $\theta$ based on the output of the predictive recursion.   

\begin{PRalgo}
Start with an initial estimate $\Psi_0$ with $\nu$-density $\pi_0\langle0\rangle + (1-\pi)\psi_0$ and a sequence of weights $w_1,\ldots,w_n \in (0,1)$.  For $i=1,\ldots,n$ compute 
\begin{align}
f_{i - 1,\theta}(Z_i) & = \int p(Z_i \mid \theta,u) \, \Psi_{i - 1}(du), \notag\\
\Psi_i(du) & = (1-w_i) \Psi_{i-1}(du) + w_i p(Z_i \mid \theta,u) \Psi_{i-1}(du) / f_{i - 1, \theta}(Z_i) \label{eq:recursion}.
\end{align}
Produce $\Psi_n$ as an estimate of $\Psi$ and $L_n(\theta) = \prod_{i = 1}^n f_{i - 1, \theta}(Z_i)$ as a marginal likelihood function for $\theta$. 
\end{PRalgo}

\citet{mt-prml} point out several justifications for labeling $L_n(\theta)$ as a likelihood function of $\theta$. For $n = 1$, $L_1(\theta)$ equals the marginal likelihood function of $\theta$, integrating out $\Psi$ under the Bayesian specification $\Psi \sim {\sf DP}(\alpha, \Psi_0)$, the Dirichlet process distribution with precision $\alpha = (1-w_1)/w_1$ and base measure $\Psi_0$.  For $n > 1$, this correspondence is not exact, but $L_n(\theta)$ can be viewed as a filtering approximation of the corresponding Dirichlet process marginal likelihood function.  Additionally, $L_n(\theta)$ features an asymptotic concentration property commonly enjoyed by likelihood functions for independent and identically distributed data models \citep{wald1949}.  Specifically, for large $n$, with $Z_1, \ldots, Z_n$ independently drawn from a common density $f^\star$, $\log L_n(\theta) \approx -nK^\star(\theta)$, where $K^\star(\theta)$ equals the minimum Kullback--Leibler divergence between $f^\star$ and densities $f$ of the form \eqref{eq:mixture} with $\Psi$ ranging over the set $\PP$ and all its weak limits points.

\section{Regularization and PRtest}
\label{S:reg}

We employ a regularized version of the predictive recursion methodology to estimate $(\theta,\Psi)$ for our two groups model.  The regularization is motivated by a hierarchical Bayes formulation of \eqref{eq:mixture} with $\Psi \sim {\sf DP}(\alpha, \Psi_0)$ where hyper-prior distributions are specified on the model parameters $\mu, \sigma, \tau$ and $\Psi_0$.  We take the $\nu$-density of $\Psi_0$ to be $\pi_0\langle0\rangle + (1-\pi_0)\psi_0$ with a fixed choice of $\psi_0(u) \propto u^2$.  Among the remaining parameters, $\sigma \in (0, \infty)$, $\tau \in (1, \infty)$ and $\pi_0 \in (0, 1)$ are taken to be independent with $\log \sigma \sim \mathsf{N}(0, 0.25^2)$, $\log(\tau - 1) \sim {\sf N}(0, 1)$ and $\pi_0 \sim {\sf Beta}(22.7, 1)$.  Given $\sigma$ and the other parameters, $\mu$ is assigned the conditional prior distribution ${\sf N}(0, \sigma^2 / 400)$. 

In our experience, $\sigma$ in the range $[0.5,2.0]$ is typical, and the log-normal prior puts nearly all of its mass there.  Other priors for $\sigma$ may also be considered, such as a conjugate scaled inverse-chi distribution.  The restriction $\tau > 1$ ensures that the non-null density $f_1$ is considerably wider than $f_0$, and the normal prior for $\log(\tau-1)$ supports a large set of values in this range.  The 22.7 in the beta prior for $\pi_0$, also used by \citet{bogdan}, assigns about 90\% of its mass to the interval $[0.9, 1]$, reflecting the belief that the null proportion $\pi$ is likely to be large.  Finally, the prior for $\mu$ is scaled to the choice of $\sigma$ and highly concentrated around the origin, reflecting the belief that the z-scores should have mean close to zero.  Finer tuning of this default prior for specific problems is straightforward.  

For a predictive recursion analog of this hierarchical Bayesian model, we interpret the predictive recursion likelihood as a function of both $\theta = (\mu, \sigma, \tau)$ and $\pi_0$. Writing this likelihood as $L_n(\mu, \sigma, \tau, \pi_0)$ and letting $g(\mu, \sigma, \tau, \pi_0)$ denote the joint prior density function on these parameters, a regularized version of the predictive recursion marginal log-likelihood function can be written as
\begin{equation}
\tilde\ell_n(\mu, \sigma, \tau, \pi_0) = \log L_n(\mu, \sigma, \tau, \pi_0) + \log g(\mu, \sigma, \tau, \pi_0).
\label{eq:regular}
\end{equation}
Estimates of these parameters are obtained by maximizing $\tilde \ell_n = \tilde \ell_n(\mu, \sigma, \tau, \pi_0)$. Once these estimates are obtained, predictive recursion is run one last time with the estimated values of these parameters to produce an estimate of $F$, i.e., of $\pi$ and of $\psi$ in \eqref{eq:two-groups} and \eqref{eq:alternative}, respectively. In our implementations, maximization of $\ell_n$ is done by the gradient-based Broyden-Fletcher-Goldfarb-Shanno (BFGS) optimization method. In Appendix~B we provide a variation on the PR algorithm that produces the gradient of $\log L_n$ as a by-product.  

The predictive recursion methodology depends on two additional factors, namely, the choice of weights $w_1,\ldots,w_n$ and the order in which the z-scores are processed by the algorithm.  \citet{mt-rate} provide an upper bound on the rate of convergence for PR estimates of the mixture $f$ when the weights are of the form $w_i = (i+1)^{-\gamma}$, $\gamma \in (2/3, 1]$.  Our choice $w_i = (i+1)^{-0.67}$ is close to the limit $\gamma = 2 / 3$ where the upper bound is optimal.  The recursive nature of the algorithm induces dependence on the order in which the $Z_i$ values are visited.  We reduce this dependence by replacing $\tilde \ell_n$ with its average over a number of random permutations of the data sequence.  Averaging over permutations increases the overall computation time, but adds stability to parameter estimation \citep{tmg}. In our experience, averaging over 10 random permutations is sufficient to stabilize the estimates of $\theta$, and the additional computation time required is negligible. To reduce variability due to random permutation, we keep the set of permutations fixed over the process of maximizing $\tilde \ell_n$.   

For multiple testing, we consider the local false discovery rate \citep{efron2004}, given by 
\[ \fdr(z) = \pi f_0(z) / f(z), \]
which represents the posterior probability that a case with z-score $Z=z$ is null.  \citet{suncai2007} argue that the local false discovery rate is the fundamental quantity for multiple testing.  Once regularized PR estimation of $(\mu,\sigma,\tau,\pi,\psi)$ is completed, a plug-in estimate $\fdrhat$ of $\fdr$ is readily available, and PRtest is implemented by thresholding $\fdrhat$; that is, we declare case $i$ as non-null if $\fdrhat(Z_i) < r$ for some specified threshold $r \in (0,1)$.  According to Efron, this multiple testing rule will control the Benjamini--Hochberg false discovery rate at level $r$.  In our examples we take $r=0.1$.  This choice, used by \citet{suncai2007}, is somewhat subjective, but sits between the choice $r=0.2$ of \citet{efron2008} and \citet{strimmer2008} and the choice $r=0.05$ of \citet{jincai2007} and others.

\section{Simulations}
\label{S:simulations}

Here we investigate the performance of PRtest in several large-scale simulations where we can compare the results with the benchmark Bayes oracle test.  The results will also be compared to those obtained from the Fourier-based method of \citet{jincai2007} and the mixfdr method of \citet{muralidharan2009}.

For $Z_1,\ldots,Z_n$, we assume independence and take the null density as $f_0(z) = {\sf N}(z \mid \mu,\sigma^2)$.  Here we fix $n=1000$, $\mu=0$, and $\sigma = 1$.  Four choices of $f_1$ are considered: 
\begin{itemize}
\item[C1:] $f_1(z) = {\sf N}(z \mid 0, \sigma^2 + \omega^2)$.  Taking $\omega^2 = 13 \approx 2\sigma^2\log n$ ensures the non-null z-scores are ``detectable'' \citep{donohojohnstone1994b}.  But, in our experience, the range of z-scores one finds in real data analysis is consistent with smaller signals, so we take $\omega^2 = 4$.  
\item[C2:] $f_1(z) = 0.5 \int_2^4 {\sf N}(z \mid u,\sigma^2) \,du$.  This choice, used by \citet{muralidharan2009} and \citet{johnstonesilverman2004}, exhibits asymmetry and has only slightly heavier tails than the null.  
\item[C3:] $f_1(z) = 0.67 \, {\sf N}(z \mid -3,2) + 0.33\, {\sf N}(z \mid3,2)$.  This one is asymmetric and a large portion of its mass is concentrated away from the origin.  
\item[C4:] $f_1(z) = 0.25 \int_{[-4,-2] \cup [2,4]} {\sf N}(z \mid u,\sigma^2) \,du$.  This is a symmetrized version of C2.  A key feature of this choice is that the unobserved signals are bounded away from zero.  
\end{itemize}
For each of the four choices of $f_1$, we consider six choices of $\pi$ ranging from 0.75 to 0.99, forming a total of 24 simulations settings. Each setting is replicated 500 times and the results are reported below.  Our implementation of PR uses weights $w_i = (i+1)^{-0.67}$ and the regularized likelihood $\tilde\ell_n$ is averaged over 10 permutations of the data sequence.

Table~\ref{tab:pi-est} summarizes the estimates of the null parameters $\pi$ for each simulation setting.  Estimates of $(\mu,\sigma)$ are similarly accurate across methods, models, and sparsity, so these results are omitted.  From the table we find that the maximum PR marginal likelihood estimates are the most adaptive across the range of $\pi$ values, specifically for choices C2--C4.  Of particular interest is PRtest's strong performance in the two most practically realistic cases, namely C3 and C4, which have smooth non-null densities with modes on both the left and right side of zero.  Also the average computation time for PRtest is roughly 3 seconds, which compares favorably with that for Jin--Cai ($0.7$ seconds) and mixfdr ($0.5$ seconds).  

\begin{table}
\begin{center}
\begin{tabular}{cc|ccc}
\hline
$f_1$ & $\pi$ & Jin--Cai & mixfdr & PRtest \\
\hline
C1 & 0.75 & 0.928 (0.019) & 0.957 (0.009) & 0.918 (0.017) \\
& 0.80 & 0.929 (0.019) & 0.965 (0.007) & 0.930 (0.016) \\
& 0.85 & 0.934 (0.018) & 0.971 (0.006) & 0.942 (0.014) \\
& 0.90 & 0.945 (0.015) & 0.980 (0.005) & 0.960 (0.014) \\
& 0.95 & 0.961 (0.011) & 0.989 (0.003) & 0.980 (0.010) \\
& 0.99 & 0.978 (0.005) & 0.995 (0.001) & 0.995 (0.003) \\
\hline
C2 & 0.75 & 0.905 (0.015) & 0.827 (0.016) & 0.761 (0.017) \\
& 0.80 & 0.874 (0.019) & 0.860 (0.012) & 0.804 (0.014) \\
& 0.85 & 0.860 (0.023) & 0.894 (0.009) & 0.851 (0.013) \\
& 0.90 & 0.869 (0.028) & 0.927 (0.007) & 0.896 (0.010) \\
& 0.95 & 0.926 (0.017) & 0.962 (0.005) & 0.940 (0.009) \\
& 0.99 & 0.984 (0.007) & 0.991 (0.003) & 0.980 (0.008) \\
\hline 
C3 & 0.75 & 0.909 (0.013) & 0.857 (0.017) & 0.788 (0.016) \\
& 0.80 & 0.886 (0.015) & 0.881 (0.013) & 0.828 (0.015) \\
& 0.85 & 0.871 (0.021) & 0.909 (0.011) & 0.867 (0.014) \\
& 0.90 & 0.886 (0.020) & 0.937 (0.008) & 0.903 (0.014) \\
& 0.95 & 0.935 (0.012) & 0.967 (0.005) & 0.937 (0.013) \\
& 0.99 & 0.980 (0.004) & 0.991 (0.003) & 0.982 (0.010) \\
\hline
C4 & 0.75 & 0.951 (0.007) & 0.886 (0.035) & 0.784 (0.066) \\
& 0.80 & 0.934 (0.010) & 0.897 (0.015) & 0.814 (0.021) \\
& 0.85 & 0.920 (0.015) & 0.920 (0.010) & 0.862 (0.018) \\
& 0.90 & 0.908 (0.025) & 0.948 (0.007) & 0.901 (0.013) \\
& 0.95 & 0.929 (0.017) & 0.975 (0.005) & 0.943 (0.012) \\
& 0.99 & 0.980 (0.007) & 0.995 (0.002) & 0.992 (0.005) \\
\hline
\end{tabular}
\end{center}
\caption{Mean (standard deviation) of the 500 estimates of $\pi$ for the method of \citet{jincai2007}, the mixfdr method of \citet{muralidharan2009}, and PRtest for the four alternative densities ($f_1$'s) described in Section~\ref{S:simulations}.}
\label{tab:pi-est}
\end{table}

\begin{figure}
\begin{center}
\scalebox{0.8}{\includegraphics{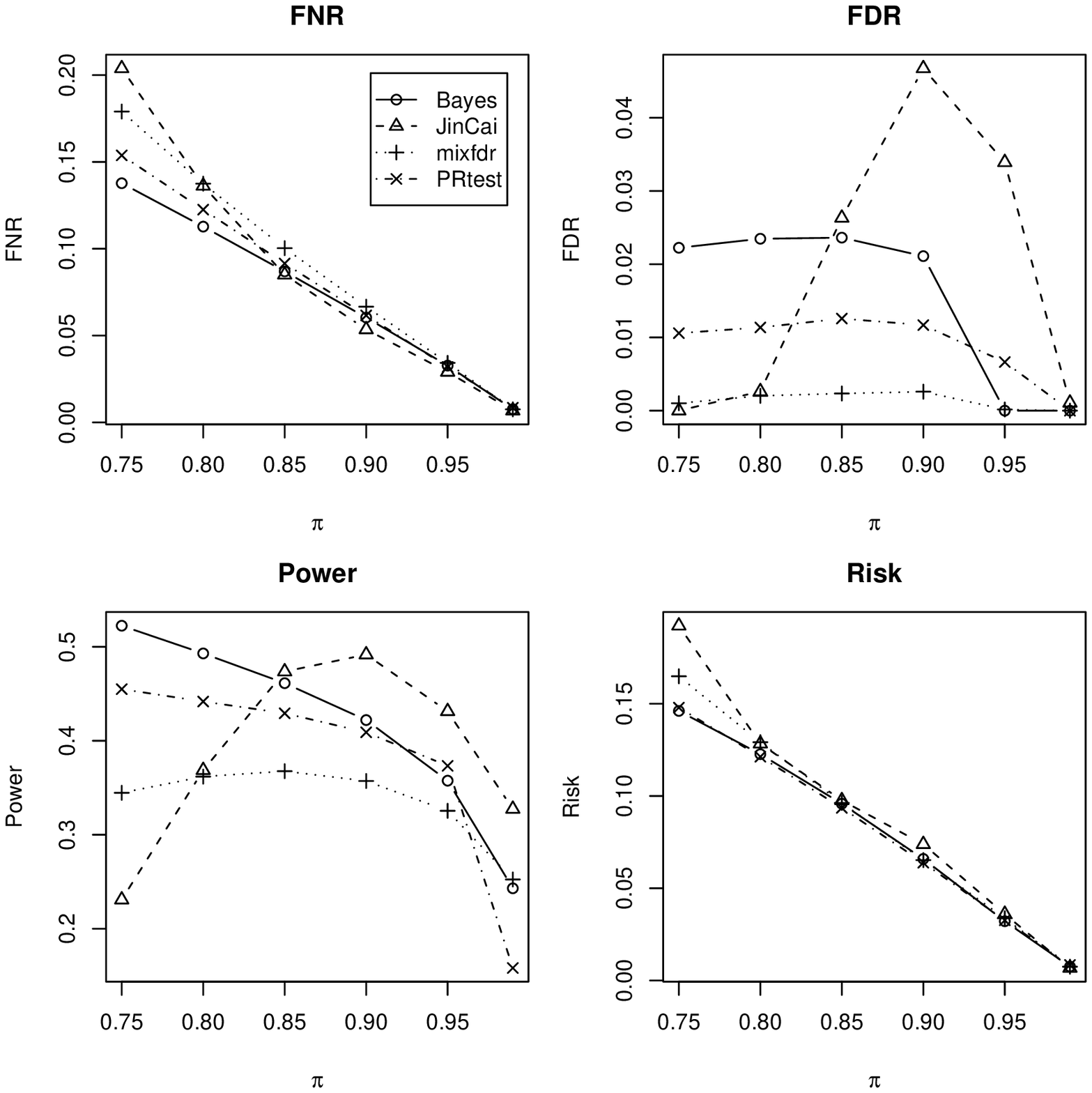}} 
\caption{Plots of the false non-discovery rate (FNR, top left), false discovery rate (FDR, top right), power (bottom left), and Bayes risk (bottom right) against $\pi$ for the selected testing procedures in the C3 simulation setting described in Section~\ref{S:simulations}. }
\label{fig:testing}
\end{center}
\end{figure}

Next we compare the performance of the selected methods based on false non-discovery rate, false discovery rate, power, and Bayes risk.  We limit this discussion to non-null choice C3; the results for the other models are similar.  Figure~\ref{fig:testing} plots these quantities as functions of $\pi$ for the selected methods and the Bayes oracle procedure; the Bayes oracle is the rule based on thresholding the \emph{true} fdr at level 0.1.  The general message is that PRtest is competitive with the other tests in all aspects across a range of sparsity levels.  In particular, the four tests are similar in terms of false non-discovery rate, particularly for large $\pi$, but PRtest is better than mixfdr and Jin--Cai for relatively small $\pi$.  Also, each of the four tests have relatively small false discovery rates, although the Jin--Cai method has a somewhat unexpected spike, which explains its higher power for large $\pi$ values.  Theoretically, the Bayes oracle test has the smallest Bayes risk uniformly over $\pi$, but the PRtest risk sits very close over the entire range of $\pi$.  This observation suggests that PRtest may be asymptotically optimal in the sense of \citet{bcfg2010}.  

\section{Examples}
\label{S:examples}

\subsection{Validation with spike-in data}
\label{SS:spike}

\begin{figure}
\begin{center}
\subfigure[PRtest fit]{\scalebox{0.71}{\includegraphics{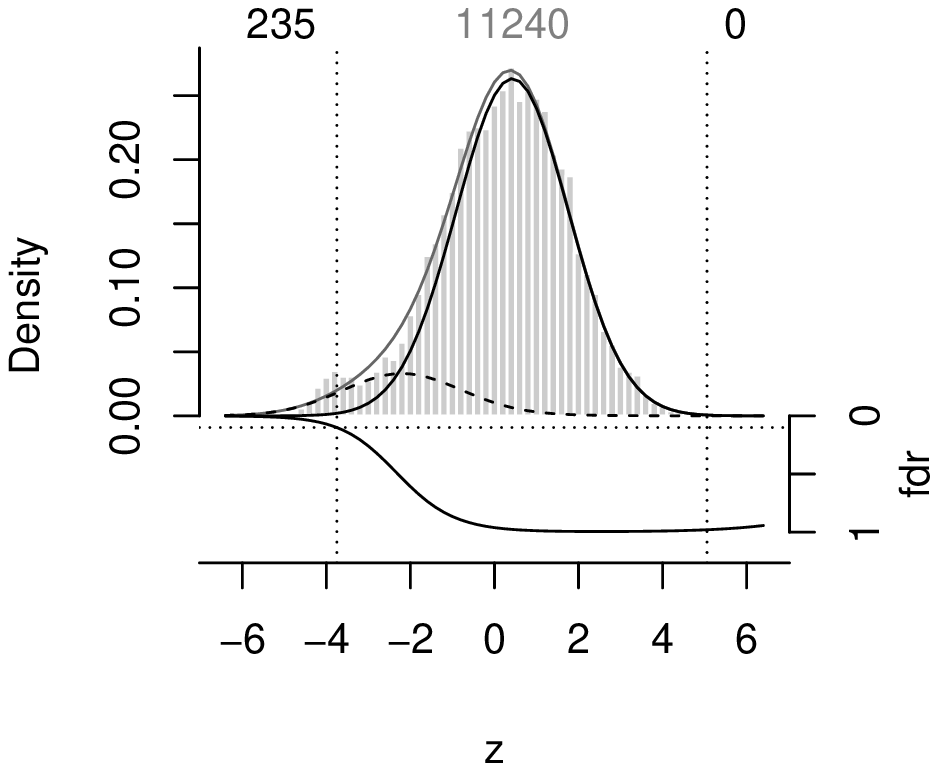}}} 
\subfigure[Oracle fit]{\scalebox{0.71}{\includegraphics{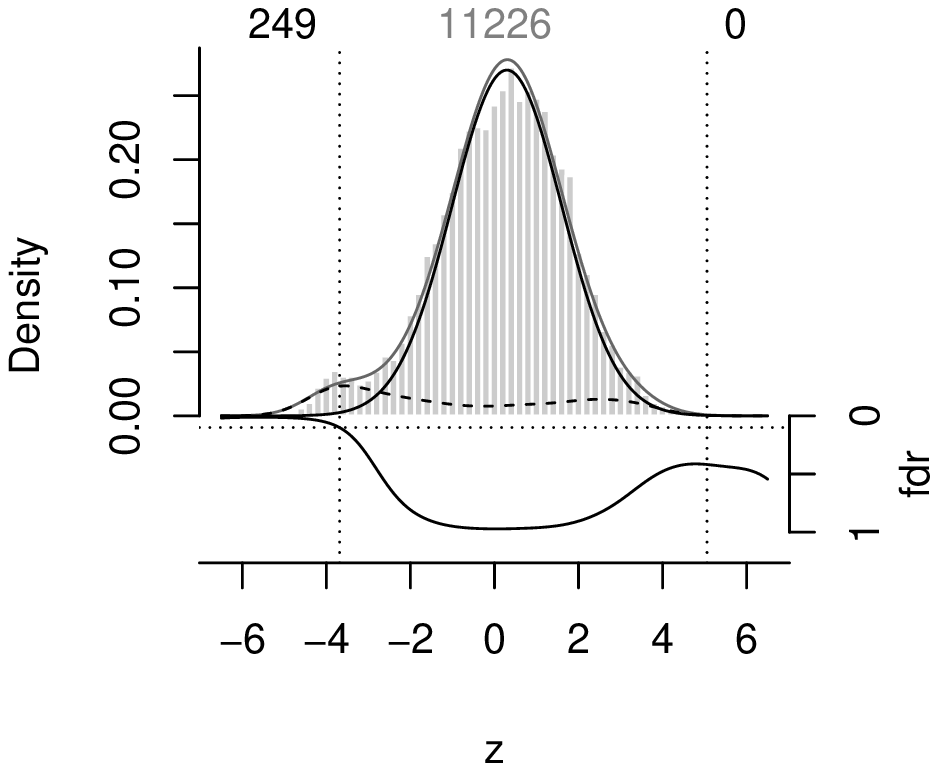}}}
\caption{Histogram of the z-scores for the golden-spike data in Section~\ref{SS:spike}, along with fits of the two-groups model using (a) PRtest and (b) the Oracle described in the text.  In each plot, overlays are $\pi f_0$ (solid black line), $(1-\pi)f_1$ (dashed black line), and $f$ (solid gray line).  The estimated fdr and the 0.1 threshold are shown on the negative scale. Numerical values on the top left and right indicate the number of genes flagged as down- and up-regulated, respectively, by the fdr thresholding rule.}
\label{fig:spike}
\end{center}
\end{figure}

An interesting \emph{spike-in} dataset was built by \citet{choe2005}.  The dataset itself is artificial---so the set of interesting genes is known---but their careful construction gives it some features of a real control-versus-treatment microarray study.  We consider a subset of this data (available in the R package {\tt st}) consisting of 11,475 genes, of which 1,331 are differentially expressed.  Z-scores are obtained by taking a Gaussian transform of the standard two-sample t-test statistics.  Figure~\ref{fig:spike}(a) shows histogram of the observed z-scores, along with the PRtest fit of the two-groups mixture model.  The estimated density clearly fits the data very well, and the fdr thresholding method flags 235 genes as down-regulated.  For comparison, Figure~\ref{fig:spike}(b) reports an oracle fit of the two-groups model, where $\pi$ is estimated as the known proportion of differentially expressed genes, $(\mu,\sigma)$ are estimated by maximum likelihood based on the null z-scores, and $f_1$ is estimated by a standard Gaussian kernel estimate based on the non-null z-scores; Table~\ref{tab:spike} reports the parameter estimates.  This oracle procedure is, in some sense, the best fdr thresholding procedure one can hope for, and it flags 249 genes as down-regulated.  

For further comparison, we applied the methods of Efron, Jin and Cai, and Muralidharan and the results are summarized in the top panel of Table~\ref{tab:spike}.  PRtest and the oracle perform similarly in every respect, while the other methods are substantially different.  Only the Jin--Cai method is able to pick out a reasonable set of interesting genes, a bit larger than the sets identified by the oracle and PRtest.  However, these additional discoveries result in a 50\% increase in false discovery rate.  

\begin{table}
\begin{center}
\begin{tabular}{ccccccccccc}
& & & & & & \multicolumn{2}{c}{Number of genes} & & & \\
\cline{7-8}
Method & & $\mu$ & $\sigma$ & $\pi$ & & Left & Right & & FDR & FNR \\
\hline
Efron & & 0.33 & 1.50 & 0.99 & & 2 & 0 & & 0\% & 12\%\\
Jin--Cai & & 0.77 & 1.45 & 0.91 & & 306 & 0 & & 3\% & 9\%\\
mixfdr & & 0.28 & 1.45 & 0.97 & & 8 & 0 & & 0\% & 12\%\\
PRtest & & 0.42 & 1.34 & 0.88 & & 235 & 0 & & 2\% & 10\%\\
{\it Oracle} & & 0.30 & 1.31 & 0.88 & & 249 & 0 & & 2\% & 10\%\\
\hline
\end{tabular}
\end{center}
\caption{Results for the spike-in dataset considered in Section~\ref{SS:spike}.  FDR and FNR denote the false discovery and false non-discovery rates, respectively.  Also, the ``Oracle'' method, as described in the text, uses the information about which genes are differentially expressed to estimate fdr.}
\label{tab:spike}
\end{table}

\subsection{Application to real data}
\label{SS:real}

\begin{figure}
\begin{center}
\subfigure[Leukemia]{\scalebox{0.72}{\includegraphics{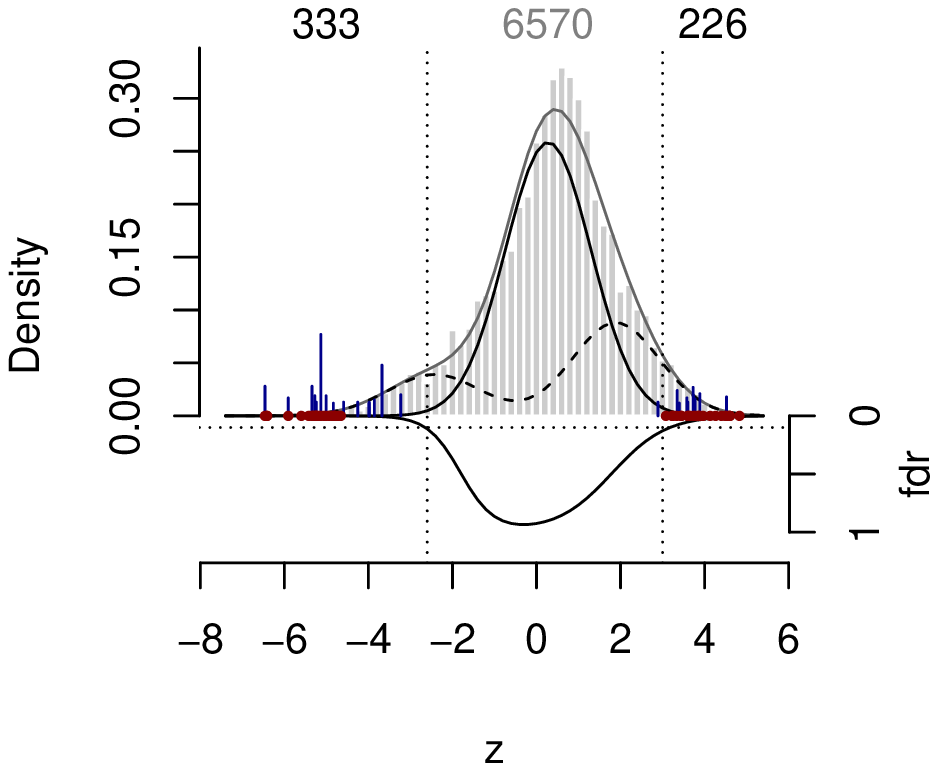}}} 
\subfigure[Breast cancer]{\scalebox{0.72}{\includegraphics{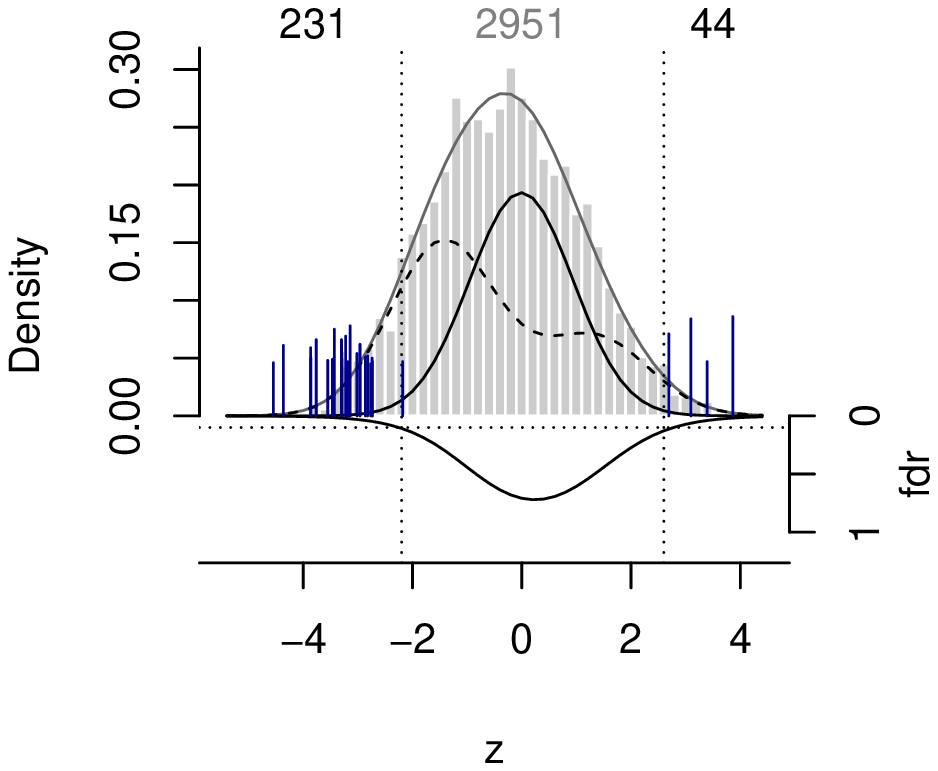}} }
\caption{PRtest's fit to z-scores values from leukemia and breast cancer microarray data. Overlaid on the z-score histogram are the estimates of $\pi f_0$ (solid, black), $(1-\pi)f_1$ (dashed, black) and $f = \pi f_0 + (1-\pi)f_1$ (solid, gray). The estimated fdr curve is shown on the negative scale, with the cut-off of 0.1 marked by the horizontal grey, dashed line. The 27 genes identified by \citet{leeplus2003} in each data set are marked with a blue bar at their z-score, with the height of the bar indicating the posterior probability of being included in the classification model. For the leukemia data, red dots are placed on the z-scores of the 50 genes identified in the original study by \citet{golub}.}
\label{fig:real}
\end{center}
\end{figure}

We applied PRtest, along with the methods of Efron, Jin and Cai, and Muralidharan, to the two microarray gene expression datasets mentioned in Section~\ref{S:intro}: the leukemia study by \citet{golub} and the hereditary breast cancer study by \citet{hedenfalk2001}.  The parameter estimates and gene classifications are summarized in Table~\ref{tab:real}.  In both datasets, PRtest estimates $\pi$ to be relatively small and identifies a number of interesting genes, while the others identify none; see Figure~\ref{fig:real}.  PRtest's findings in these two datasets are corroborated by the results of \citet{leeplus2003} who learn a treatment classifier from gene expression levels and validate it by accurately classifying samples from an independent test set.  That is, the set of interesting genes identified by PRtest substantially overlaps with the set of genes \citet{leeplus2003} flag as important constituents of their classifier; these are also displayed in Figure~\ref{fig:real}.  For the breast cancer study, some of the genes identified by PRtest and \citet{leeplus2003}, such as keratin~8, TOB~1, and phosphofructokinase platelet, have known biological connections to breast cancer mutations \citep[][p.~93]{leeplus2003}.  The fact that the gene expression levels lead to a well-validated classifier suggests that some genes must be differentially expressed.  In this light, it is surprising that the methods of Efron, Jin and Cai, and Muralidharan fail to identify a single interesting gene.

\begin{table}
\begin{center}
\begin{tabular}{cccccccccccc}
 & & & & & & & \multicolumn{2}{c}{Number of genes} \\
\cline{8-9}
Data & Method & & $\mu$ & $\sigma$ & $\pi$ & & Left & Right \\
\hline
Leukemia & Efron & & 0.57 & 1.18 & 0.88 & & 276 & 0 \\
& Jin--Cai & & 0.95 & 1.30 & 0.91 & & 291 & 0 \\
& mixfdr & & 0.56 & 1.35 & 0.96 & & 71 & 0 \\
& PRtest & & 0.23 & 1.04 & 0.63 & & 333 & 226 \\
\hline
BRCA & Efron & & $-0.33$ & 1.45 & 1.00 & & 0 & 0 \\
& Jin--Cai & & $-0.42$ & 1.44 & 1.00 & & 0 & 0 \\
& mixfdr & & $-0.31$ & 1.38 & 0.99 & & 0 & 0 \\
& PRtest & & $-0.01$ & 1.04 & 0.45 & & 231 & 44 \\
\hline 
\end{tabular}
\end{center}
\caption{Results for the two real microarray datasets considered in Section~\ref{SS:real}.}
\label{tab:real}
\end{table}

\ifthenelse{1=1}{}{
\begin{table}
\centering
\begin{tabular}{r|ccc|ccc}
& \multicolumn{3}{c}{\it Leukemia} & \multicolumn{3}{|c}{\it Breast Cancer}\\
 & $\hat \pi$ & Left & Right & $\hat \pi$ & Left & Right \\
\hline 
Efron & 0.88 & 276 & 0 &  1.00 & 0 & 0\\
Jin--Cai & 0.91 & 291 & 0 & 1.00 & 0 & 0\\
mixfdr & 0.96 & 71 & 0 & 0.99 & 0 & 0\\
PRtest & 0.63 & 333 & 226 & 0.45 & 231 & 44 \\
\hline
\end{tabular}
\caption{A comparison of PRtest with the methods by \citet{efron2004}, \citet{jincai2007} and \citet{muralidharan2009} for leukemia and breast cancer gene expression data. Methods are compared based on the null proportion estimate and the number of significant genes identified in the left and right tail  of the z-scores histogram in Figure~\ref{fig:real}.}
\label{tab:real}
\end{table}
}

\section{Discussion}
\label{S:discuss}

This paper provides a new and identifiable semiparametric formulation of the two-groups model and a computationally efficient algorithm to estimate the model parameters.  This naturally leads to a nonparametric empirical Bayes multiple testing rule based on thresholding the estimated local false discovery rate.  In simulations we find that PRtest is comparable to existing methods, including the Bayes oracle.  What is particularly interesting is that the PRtest results differ substantially from those of existing methods in the examples of Section~\ref{S:examples}, and we argue that our findings are, in fact, more believable.  

We have chosen to focus only on the case where the null z-scores are normally distributed, though the theory and methods presented here work for other well behaved parametric families. Normality of null z-scores is indeed a strong structural assumption, but identification of the null from the non-null requires strong parametric shape restrictions on one of the two components.  Assuming a normal null component is natural because, theoretically, the null z-scores should have a standard normal distribution. This is similar to p-value-based methods where the null p-values are assumed to be uniform. A purely statistical verification of this kind of assumption seems quite challenging. One could possibly gain insight on this issue through biological experiments consisting entirely of null cases.

We have justified the continuous location mixture formulation of $f_1$ in \eqref{eq:alternative} on two grounds: first, it makes the model parameters identifiable and, second, it conforms to the accepted notion that the alternative is more likely than the null to produce z-scores of large magnitude.  This latter property is also satisfied by a discrete mixture $f_1 = \sum_{j = 1}^J \pi_j {\sf N}(\mu + \tau\sigma u_j, \sigma^2)$, for which the identifiability condition does not hold.  But with the regularization to encourage selection of $f_0$ centered near zero, and the ability of a flexible continuous mixture to approximate a discrete one, PRtest might still perform well in this difficult situation.  Our limited simulations seem to indicate that this is true.  The case where $f_1$ is not wider than $f_0$ also yields a coherent statistical simulation model, but we argue that it corresponds to a biologically untenable abstraction.  Indeed, the multiple testing framework accepts the z-scores as scores whose magnitudes (possibly after a small shift of origin) give an ordering of how interesting the cases are relative to each other.  The question is to decide how interesting a case must be in order to be labeled as non-null.  Accepting the relative ordering is equivalent to accepting that $f_1$ must be wider than $f_0$.

\ifthenelse{1=1}{}{
In this section we discuss several important issues regarding our model assumptions in Section~\ref{S:model} and the conclusions reached in Sections~\ref{S:simulations} and \ref{S:examples}.

We have chosen to focus only on the case where the null z-scores are normally distributed, though the theory and methods presented here work for other well behaved parametric families. Normality of null z-scores is indeed a strong structural assumption, but to be able to identify the null from the non-null it is important to impose strong parametric shape restrictions on one of the two components. Assuming a normal null component is natural because, theoretically, the null z-scores are supposed to have a standard normal distribution. This is similar to p-value-based methods where the null p-values are assumed to be uniform. A purely statistical verification of this kind of assumptions seems quite challenging. One could possibly gain insight on this issue through careful biological experiments consisting entirely of null cases.

We have justified the continuous location mixture formulation of $f_1$ in \eqref{eq:alternative} on two grounds: first, it makes the model parameters identifiable and, second, it conforms to the accepted notion that the alternative is more likely than the null to produce z-scores of large magnitude.  This latter property is also satisfied by a discrete mixture $f_1 = \sum_{j = 1}^J \pi_j \mathrm{N}(\mu + \tau\sigma u_j, \sigma^2)$, for which the identifiability condition does not hold.  But with the regularization to encourage selection of $f_0$ centered near zero, and the ability of a flexible continuous mixture to approximate a discrete one, there is still hope for PRtest in this difficult situation.  For a concrete example, consider model C3 of Section~\ref{S:simulations} but with variances equal to 1 instead of 2.  In 50 random data sets from this model, with $\pi = 0.9$, the average PR estimates of $(\mu, \sigma, \pi)$ are $(-0.02, 1.02, 0.91)$, with standard deviations $(0.02, 0.02, 0.01)$.  Not only are these estimates more accurate than those obtained by both the Jin--Cai and mixfdr methods, they suggest that PRtest is able to identify the null, even in this non-identifiable setting.

An interesting philosophical question is what happens if $f_1$ is not wider than $f_0$.  While this presents a coherent statistical simulation model, it corresponds to a biologically untenable abstraction.  The multiple testing framework accepts the z-scores as scores whose magnitudes (possibly after a small shift of origin) give an ordering of how interesting the cases are relative to each other.  The question is to decide how interesting a case must be in order to be labeled as non-null.  Accepting the relative ordering is equivalent to accepting that the alternative density must be wider than the null.
 
For the leukemia and the hereditary breast cancer studies, where existing multiple testing methods fail to identify any genes, we have defended PRtest's discovery of several interesting genes by pointing out similar discoveries by high-dimensional classification-based methods. Our position is that if a set of genes can be identified to produce significantly accurate classification of the two treatment conditions, then it is preposterous for any statistical study to declare that there is no interesting genes in the collection. Of course, classification accuracy needs to be beyond a reasonable doubt of statistical serendipity. Use of strong shrinkage type regularization and test case validation help dispel such doubts; both of these have been used in the studies by \citet{leeplus2003}.

We also take the stance that while classification-based methods can be trusted to decide whether there is a set of interesting genes, they cannot be trusted to reveal the set of all interesting genes, or even a set of most interesting genes. This is only due to our limited computing resources which force us to use stochastic exploration of how interesting various subsets of genes are. Employing an exhaustive calibration of all $2^n$ subsets is clearly infeasible. This is where multiple testing can be more informative as it lines up all genes based on a scalar score of ``interestingness'' and restricts exploration of interesting subsets through simple splicing of this score vector. Even with stochastic exploration, classification-based methods are fairly computationally intensive, whereas fast computation is the hallmark of multiple testing.
}

\section*{Software}

R software to implement the proposed PRtest methodology can be found at S.~Tokdar's website, \url{http://www.stat.duke.edu/~st118/Software}.

\section*{Acknowledgments}

The authors are grateful to the Editor, Associate Editor, and two anonymous referees for their insightful comments and suggestions, and to Professor J.~K.~Ghosh for helpful discussions.  A portion of this work was completed while R.~Martin was with the Department of Mathematical Sciences, Indiana University--Purdue University Indianapolis.

\appendix

\section{Proof of Theorem~1}

Here we prove a more general version of Theorem~1 in the main text.  Let $p(z)$ be a probability density function on $\RR$, symmetric about zero.  Furthermore, assume $p$ is supersmooth in the sense of \citet{fan1991}; see \eqref{eq:supersmooth} below.  In the main text, we took $p(z)$ to be a ${\sf N}(0,1)$ kernel but, e.g., a Student-t kernel with known degrees of freedom would also satisfy these conditions.  

For the particular choice of $p$, define the density-valued mapping 
\begin{equation*}
\begin{split}
M(\mu,\sigma,\tau, \pi, \psi)(z) & = \pi \sigma^{-1}p\bigl( (z-\mu)/\sigma \bigr) \\
& \qquad + (1-\pi) \int \sigma^{-1} p\bigl( (z-\mu-\tau u)/\sigma \bigl) \psi(u) \,du. 
\end{split}
\end{equation*}
To prove that $(\mu,\sigma,\tau,\pi,\psi)$ are identifiable, we need to show that $M$ is a one-to-one function.  Therefore, we start by assuming $M(\mu_1,\sigma_1,\tau_1,\pi_1,\psi_1) = M(\mu_2,\sigma_2,\tau_2,\pi_2,\psi_2)$.  Let $p^*(t)$ and $\psi_k^*(t)$ denote the characteristic functions of $p(z)$ and $\psi_k(z)$, respectively, for $k=1,2$.  Then we must have 
\begin{align}
\exp(it\mu_1) p^*(\sigma_1 t) \bigl\{ &\pi_1 + (1-\pi_1) \psi_1^*(\sigma_1t/\tau_1) \bigr\} \notag \\ & = \exp(it\mu_2) p^*(\sigma_2 t) \bigl\{ \pi_2 + (1-\pi_2) \psi_2^*(\sigma_2 t / \tau_2) \bigr\} \label{eq:chfn}
\end{align}
for every $t \in \RR$, where $i$ is the imaginary unit.  Recall that the Riemann--Lebesgue lemma \citep[e.g.,][Theorem~26.1]{billingsley} says, for $k=1,2$, 
\begin{equation}
\label{eq:chfn-limit}
\psi_k^*(t) \to 0 \quad \text{as} \quad t \to \pm \infty.
\end{equation}
Now, suppose $\sigma_1 > \sigma_2$ and assume, without loss of generality, that $\mu_2 > 0$.  Choose a sequence $\{t_s\} \subset \RR$ such that $t_s \to \infty$ and  $\exp(it_s \mu_2) \equiv 1$.  Then, for large enough $s$, \eqref{eq:chfn-limit} implies that $\pi_2 + (1-\pi_2)\psi_2^*(\sigma_2 t_s/\tau_2) \neq 0$.  On rearranging the terms in \eqref{eq:chfn} we get 
\begin{equation}
\label{eq:chfn-ratio}
\frac{p^*(\sigma_2t_s)}{p^*(\sigma_1t_s)} = \frac{\exp(i t_s \mu_1) \bigl\{ \pi_1 + (1-\pi_1)\psi_1^*(\sigma_1 t_s/\tau_1) \bigr\}}{\pi_2 + (1-\pi_2) \psi_2^*(\sigma_2 t_s/\tau_2)}. 
\end{equation}
We have assumed that $p$ is supersmooth \citep{fan1991}, which means that
\begin{equation}
\label{eq:supersmooth}
d_0 |t|^{\beta_0} \exp\{-|t|^\beta/\gamma\} \leq |p^*(t)| \leq d_1 |t|^{\beta_1} \exp\{-|t|^\beta/\gamma\}, 
\end{equation}
for all $t$ and for some positive constants $d_0, d_1, \beta_0, \beta_1, \beta$, and $\gamma$.  Under this assumption, the modulus of the left-hand side of \eqref{eq:chfn-ratio} satisfies
\[ \Bigl| \frac{p^*(\sigma_2t_s)}{p^*(\sigma_1t_s)} \Bigr| \geq \text{const} \times |t_s|^{\beta_1-\beta_0} \exp\{ |t_s|^\beta (\sigma_1^\beta - \sigma_2^\beta) / \gamma\}. \]
Therefore, as $s \to \infty$, the left-hand side of \eqref{eq:chfn-ratio} is unbounded while the right-hand side is bounded.  This is a contradiction, so we need $\sigma_1 \leq \sigma_2$.  But by symmetry, it follows that $\sigma_1 = \sigma_2$.  With this equality, relation \eqref{eq:chfn} easily leads to the equalities $\mu_1 = \mu_2$, $\tau_1=\tau_2$, $\pi_1 = \pi_2$ and $\psi_1 = \psi_2$, completing the proof.

\section{Gradient of the log PR marginal likelihood}

This section provides a variation on the predictive recursion (PR) algorithm that yields the gradient of the log PR marginal likelihood function, based on the development in \citep{mt-prml}.  The model under consideration here is the following:
\[ f(z) = \pi {\sf N}(z \mid \mu,\sigma^2) + \bar\pi \int {\sf N}(z \mid \mu + \tau \sigma u, \sigma^2) \psi(u) \,du, \]
where $\psi$ is an unknown mixing density supported on $[-1,1]$.  The details of the PRtest method can be found in the main text.  Here we focus only on computing the gradient of $\ell_n(\theta) = \sum_{i=1}^n \log f_{i-1,\theta}(Z_i)$, where $f_{k,\theta}(z)$ is the PR estimate of the mixture density based on $Z_1,\ldots,Z_k$ and $\theta = (\mu, \sigma, \tau, \pi_0)$, slightly different than in the main text.  

Define an unconstrained version of $\theta$, i.e., $\eta = (\mu, \log\sigma, \log(\tau-1), \logit\pi_0)$, where $\logit x = \log(\frac{x}{1-x})$.  In what follows, $\grad$ will denote a gradient with respect to $\eta$, and if $g$ is a function of a variable $u$, then $\grad g(u)$ denotes the gradient with respect to $\eta$, pointwise in $u$.  The following algorithm shows how to compute $\lambda_i = f_{i-1,\theta(\eta)}(Z_i)$ and $\grad \log \lambda_i$ for $i=1,\ldots,n$.  

\begin{enumerate}

\item Start with user-specified $\pi_0$ and $f_0$, and set 
\[ \grad \pi_0 = (0, 0, 0, \pi_0(1-\pi_0)) \quad \text{and} \quad \grad \psi_0(u) \equiv (0,0,0,0). \]

\item For $i=1,\ldots,n$, repeat the following three steps:

\begin{enumerate}

\item For the normal kernel $p(z \mid \theta,u) = {\sf N}(z \mid \mu + \sigma\tau u, \sigma^2)$, set
\[ G_0 = {\sf N}(Z_i \mid \mu, \sigma^2) \quad \text{and} \quad G_1(u) = {\sf N}(Z_i \mid \mu + \sigma \tau u, \sigma^2), \]
and analytically evaluate the gradients $\grad G_0$ and $\grad G_1(u)$: 
\begin{align*}
\grad G_0 & = (z_0/\sigma, z_0^2-1, 0, 0) \cdot G_0 \\
\grad G_1(u) & = (z_1(u)/\sigma, z_1(u) \tau u / \sigma + z_1^2(u) - 1, z_1(u) u (\tau-1), 0) \cdot G_1(u),
\end{align*}
where $z_0 = (Z_i-\mu)/\sigma$ and $z_1(u) = (Z_i-\mu-\sigma\tau u)/\sigma$.  

\item Compute
\begin{align*}
h_i & = \int G_1(u) \psi_{i-1}(u) \,du \\
\lambda_i & = \pi_{i-1} G_0 + (1-\pi_{i-1}) h_i \\
\grad \log h_i & = \frac{1}{h_i} \int \bigl\{ G_1(u) \grad \psi_{i-1}(u) + \grad G_1(u) \psi_{i-1}(u) \bigr\} \,du \\
\grad \log \lambda_i & = \frac{\grad \pi_{i-1} G_0 + \pi_{i-1} \grad G_0 + h_i \{ (1-\pi_{i-1}) \grad \log h_i - \grad \pi_{i-1} \} }{u_i} 
\end{align*}

\item Update
\begin{align*}
\pi_i & = A_0 \pi_{i-1} \\
\grad \pi_i & = A_0 \grad \pi_{i-1} + \grad A_0 \pi_{i-1} \\
\psi_i(u) & = B A_1(u) \psi_{i-1}(u)  \\
\grad \psi_i(u) & = \{ \grad B A_1(u) + B \grad A_1(u) \} \psi_{i-1}(u) + B A_1(u) \grad \psi_{i-1}(u) 
\end{align*}
where 
\begin{align*}
A_0 & = 1 + w_i (G_0 / \lambda_i - 1) \\
A_1(u) & = 1 + w_i (G_1(u) / \lambda_i - 1) \\
B & = (1-\pi_{i-1}) / (1-A_0 \pi_{i-1})
\end{align*}
and
\begin{align*}
\grad A_0 & = w_i \{ \grad G_0 - G_0 \grad \log \lambda_i \} / \lambda_i \\
\grad A_1(u) & = w_i \{ \grad G_1(u) - G_1(u) \grad \log \lambda_i \} / \lambda_i \\
\grad B & = \frac{(BA_0-1)\grad\pi_{i-1} + B \grad A_0 \pi_{i-1}}{1-A_0\pi_{i-1}} 
\end{align*}

\end{enumerate}

\item Return the log-likelihood $\sum_{i=1}^n \log \lambda_i$ and its gradient $\sum_{i=1}^n \grad \log \lambda_i$.

\end{enumerate}

\bibliographystyle{/Users/rgmartin/Research/TexStuff/asa}
\bibliography{/Users/rgmartin/Research/mybib}

\end{document}